\documentclass[preprint,aps]{article}

\begin{document}
\begin{titlepage}
\centerline{\large \bf Photoionization of Ce$^{3+}$-Ce$^{4+}$}  
\vglue .1truein
\centerline{Zhifan Chen and Alfred Z. Msezane}
\centerline{\it Center for Theoretical Studies of Physical Systems,  and
Department of Physics}
\centerline{\it Clark Atlanta University, Atlanta, Georgia 30314, U. S. A.}
\centerline{}
\centerline{}
\centerline{\bf ABSTRACT}
Photoionization of the Ce$^{3+}$ - Ce$^{4+}$ process has been studied using the random 
phase approximation 
with exchange method in the energy region 100-150 eV.  Comparison of our results   
with the recent measurement [M\"uller {\it et al}, Phys. Rev. Lett. {\bf 101}, 133001
(2008)] confirms the suppression effect of the carbon cage
in the endohedral fullerene Ce@C$_{82}^+$ photoionization. 
The reasons for the cause of the confinement 
resonance and the suppression effect are discussed. 

\end{titlepage}
\leftline{\bf 1. Introduction}
The motivation to study the photoionization of the Ce$^{3+}$ - Ce$^{4+}$ process 
is the disturbing     
discrepancy between the theoretical calculations 
and the experimental measurements of the photoionization of an endohedrally confined atom. 
The theoretical calculations [1-6] indicate the presence of strong 
confinement resonances for the endohedral fullerenes, such as Xe@C$_{60}$ [6]. 
However, the experimental results [7-11] demonstrate a great suppression of 
the photoionization cross section of an atom encapsulated in the carbon sphere,
for example the photoionization of   
Ce@C$_{82}$ [7], Pr@C$_{82}$ [8], Ce@C$_{82}^+$ [9, 10], and Dy@C$_{82}$ [11]. 

Both experiments [12, 13] and theoretical study [14] have indicated that the Ce atom in 
the endohedral fullerene Ce@C$_{82}$
is located at an 
off-centered position adjacent to the carbon cage and the encapsulated Ce atom donates 
three valence 
electrons to the carbon sphere.
The electronic state of Ce@C$_{82}$ can be formally described as Ce$^{3+}$@C$_{82}^{3-}$ [13].
The same charge state can be found for the Ce atom in the endohedral molecule Ce@C$_{82}^+$ [10].
Since the photoelectron ionized from the Ce$^{3+}$ ion of the Ce@C$_{82}^+$ molecule 
will be multiply reflected to different directions by the carbon cage, 
it is difficult from a theoretical point of view to consider the photoionization of 
an off-centered Ce$^{3+}$ ion.
However, the photoionization of the Ce$^{3+}$ ion can be calculated by using our 
recently developed random phase approximation with exchange (RPAE) method [15] if 
the necessary modifications
in the computer code are made.   
The results should further confirm the cage suppression 
effect if agreement is obtained with the 
experimental data for the Ce$^{3+}$ - Ce$^{4+}$ photoionization [10]    
or a transparent cage model can be set up for the photoelectron if it  
agrees with the photoionization cross section of the Ce@C$_{82}^+$. 

The RPAE method,
which allows for the inclusion of both intrashell and intershell correlations
has been developed recently by Chen and Msezane for atoms(ions) with an outer open-shell [15],
or with an inner open shell [16] and
has been successfully used to study the
$4d - \epsilon f$ photoionization 
of the ions Xe$^+$ [15] and I$^+$ [16] and the photoelectron angular distribution asymmetry 
parameter 
$\beta$ of the Sc 4s electron [17].   
In this calculation our RPAE codes are modified to include the  
intershell coupling between the Ce$^{3+}$ $5s-\epsilon p$, $5p-\epsilon s, d$ and
$4d-4f$ transitions. A new computer code is used to study the Ce$^{3+}$ photoionization. 

\leftline{\bf 2. Theory}

The RPAE equation and the symbols and operators in the equation 
for an atom with an outer open-shell 
is given by Eq. (1) of Ref [15].
Similar terms will be added 
for the switch of other electron pairs.
All the matrix elements have been derived and presented in the Appendices of 
Refs. [15] and [17]. The Coulomb matrix elements which are needed to evaluate the 
intershell coupling between the 
$5s-\epsilon p$, $5p-\epsilon s, \epsilon d$ and
$4d-4f$ transitions of the  
Ce$^{3+}$ ion,  
are given in the Appendix of this paper.

The ground state of the Ce$^{3+}$ ion has the configuration [Xe]4f$(^2F)$. 
Here we assume the $5d$ and $6s^2$ electrons of the Ce atom, which are located 
outside the $4f$ orbital
have been transferred to the carbon 
cage in the endohedral fullerene Ce@C$_{82}^+$.

Since we try to include the intershell couplings among the $5p-\epsilon s,\epsilon d$
,5$s-\epsilon p$, and $4f-\epsilon d, \epsilon g$ transitions , the
following combined core with discrete and continuum electron states have been included 
in the calculation.

We have a total of 21 channels from the $5p+h\nu \rightarrow\epsilon d, \epsilon s $ 
transitions 
to the states:

$4d^{10} 5s^2 5p^5 4f (^1D,\ ^1G,\ ^1F,\ ^3D,\ ^3G,\ ^3F,) \epsilon d(^2D,\ ^2F,\ ^2G)$,

$4d^{10} 5s^2 5p^5 4f (^1D,\ ^3D) \epsilon s (^2D)$,

$4d^{10} 5s^2 5p^5 4f (^1F,\ ^3F) \epsilon s (^2F)$,

$4d^{10} 5s^2 5p^5 4f (^1G,\ ^3G) \epsilon s (^2G)$,

a total of 6 channels from the $5s+h\nu \rightarrow \epsilon p$ transition to the states:

$4d^{10} 5s^2 5p^5 4f (^1F,\ ^3F) \epsilon p (^2D,\ ^2F,\ ^2G)$,

and a total of 2 channels from the $4f+h\nu \rightarrow \epsilon d, \epsilon g$ 
transitions to the states:

$4d^{10} 5s^2 5p^6  (^1S) \epsilon d (^2D)$,

$4d^{10} 5s^2 5p^6  (^1S) \epsilon g (^2G)$.

The closed shells of $1s,2s,2p,3s,3p,4s$, and $4p$ are not listed above. 
As the photoionization process is caused mainly by the autoionization of the $4f$ subshell  
our code has several special subroutines to treat the 
$4d^{10}4f +h\nu\rightarrow 4d^9 4f^2$ 
transition, 
with a total of 14 channels:      

\begin{equation}
4d^{10} 5s^2 5p^6 4f +h\nu\rightarrow 4d^9 5s^2 5p^6 (4f^2(^3H)) ( ^2F,\ ^2G)
\end{equation}
\begin{equation}
4d^{10} 5s^2 5p^6 4f +h\nu\rightarrow 4d^9 5s^2 5p^6 (4f^2(^3F)) ( ^2D,\ ^2G)
\end{equation}
\begin{equation}\noindent
4d^{10} 5s^2 5p^6 4f +h\nu\rightarrow 4d^9 5s^2 5p^6 (4f^2(^3P)) ( ^2D,\ ^2F)
\end{equation} 
\begin{equation}
4d^{10} 5s^2 5p^6 4f +h\nu\rightarrow 4d^9 5s^2 5p^6 (4f^2(^1I)) ( ^2G)
\end{equation}
\begin{equation}
4d^{10} 5s^2 5p^6 4f +h\nu\rightarrow 4d^9 5s^2 5p^6 (4f^2(^1G,\ ^1D)) ( ^2D,\ ^2F,\ ^2G)
\end{equation}
\begin{equation}\noindent
4d^{10} 5s^2 5p^6 4f +h\nu\rightarrow 4d^9 5s^2 5p^6 (4f^2(^1S)) ( ^2D)
\end{equation}

The Ce$^{3+}$ ground state and the core wave functions were obtained through the 
self-consistent 
Hartree-Fock (HF) calculation. Then the radial functions of the discrete and continuum electron
were obtained by solving the linear HF equations without self-consistency using those core 
wave functions. Each radial function has been represented by 2000 points. 
After evaluating the dipole matrix elements and the Coulomb matrix elements of 
the time-forward type and the time-backward type, the RPAE equation 
was solved to obtain the partial cross sections with a total 
of 15 $^2D$ states, 15 $^2G$ states, and 13 $^2F$ states.  
All types of matrix elements are evaluated using the equations found in the Appendices of 
Ref. [15] and Ref. [16]. Equations for the Coulomb matrix elements of the intershell
coupling between the Ce$^{3+}$ $5p-\epsilon s, \epsilon d$, $5s-\epsilon p$ 
and $4d-4f$ transitions are found in the
Appendix of this paper.

\leftline{\bf Results}

Its well known that there are two kinds of giant resonances [18]. One is 
the shape resonance in which the properties of the giant 
resonance is determined mainly by the effective potential for the $f$ electron. 
The overlaps of bound $f$ orbital with the 4$d$ 
orbital are very small.  Most of the $4d\rightarrow f$ oscillator strength is associated 
with the continuum state. 
The photoionization of the $4d-\epsilon f$ channel in the atoms Xe, Ba, and I [19] and 
ions Xe$^+$ [15] and I$^+$ [16] are the
examples of this type of giant resonance. The resonance results from a one step process: 
continuum 
enhancement due to the centrifugal-barrier shape resonance.  
The second type of giant resonance corresponds to a decaying  
discrete resonance, which results from a two-step process: photoexcitation of a 
$4d$ electron into the $4f$ subshell, followed by autoionization of the 
$4d^9 4f^{N+1}$ state where $N$ is the initial occupation number for the $4f$ electron.   
An example of this type of giant resonance is found in the photoionization of the 
rare earths elements.  The decaying discrete resonance is related to the so-called 
collapsed $f$-wave function. 
In this situation 
the inner well is deep enough to support a bound state and the $4f$ orbital "collapses" 
into the inner well.  

The photoionization of the Ce$^{3+}$ 4$d$ electron belongs to the second type of giant 
resonance. 
The mean radus of the $4f$ electrons is only 0.96 a.u., which is much smaller than that of
the $5s$ electrons (1.58 a.u.) and $5p$ electrons (1.75 a.u.). Most of the $4d\rightarrow f$
oscillator strength is associated with the $4f$ state. We performed 
two calculations: one with the processes represented by equations (1)-(6), and the 
other without 
these processes. The results show that the photoionization cross section 
without the $4d-4f$ transition is about only 10\% of the results when these  
transitions are included. It demonstrates that the photoionization cross section is 
mainly (90\%)
caused by two processes. The first process is to photoexcite the $4d$ electron to 
the $4f$ subshell, then
the autoionization of the $4f$ subshell causes the 
photoionization of the $4f, 5s,$ and $5p$ subshells.

Figures 1 gives our calculated   
photoionization cross sections for the Ce$^{3+}$-Ce$^{4+}$ transition 
in the energy region 100-150 eV.
Dotted, dashed and dash-dotted curves represent, respectively the partial cross sections 
for the $^2F$, $^2D$ and $^2G$ states. The first peak, located at 101.7 eV, is 
contributed mainly (88\%) by the partial cross section of the $^2F$ state. 
The most important photoexcited states are 

$4d^9 5s^2 5p^6 (4f^2(^1G)) (^2F)$, and 

$4d^9 5s^2 5p^6 (4f^2(^3H)) (^2F)$.

The most important final states through intershell coupling among the $5p-\epsilon d$ 
$5s-\epsilon p$ and $4d-4f$ transitions are   

$4d^{10} 5s^2 5p^5 4f (^1G) \epsilon d (^2F)$, 

$4d^{10} 5s^2 5p^5 4f (^3D) \epsilon d (^2F)$,

$4d^{10} 5s^2 5p^5 4f (^3G) \epsilon d (^2F)$, 

$4d^{10} 5s 5p^6 4f (^3F) \epsilon p (^2F)$,

$4d^{10} 5s 5p^6 4f (^1F) \epsilon p (^2F)$.

Other transitions will also contribute to the cross section but to a lesser extent. 
We note that the above 
five states contribute 86.0\% to the partial cross section of the $^2F$ state  
in the first peak at 101.7 eV.

The second peak is at 118.0 eV with a maximum cross section of 47.9 Mb.  The
$^2D$, $^2F$ and $^2G$
states contribute respectively 2.4 Mb, 3.9 Mb and 41.6 Mb to this peak. 
The transition to the 
$^2G$ state accounts for 
86.8\% of the total cross section. 
Therefore, the most important processes are first $4d$ electron excited to the
states

$4d^9 5s^2 5p^6 (4f^2(^3H))(^2G)$, and

$4d^9 5s^2 5p^6 (4f^2(^1I))(^2G)$.

Then autoionization follows to the final states 

$4d^{10} 5s^2 5p^6(^1S)\epsilon g (^2G)$,

$4d^{10} 5s^2 5p^5 4f (^3D)\epsilon d (^2G)$,

$4d^{10} 5s^2 5p^5 4f (^3F)\epsilon d (^2G)$,

$4d^{10} 5s 5p^6 4f (^3F)\epsilon p (^2G)$.

Since the $4f-\epsilon g$ transition has the cross section 20.4 Mb, which is almost half
the cross section (41.6 Mb) of the $^2G$ state, this transition is the most important 
process for the second peak.

The third peak has a maximum of 57.0 Mb and is due to the partial cross sections for 
the $^2D$, $^2F$ and $^2G$ states whose values are respectively, 23.7 Mb, 26.0 Mb, 
and 7.3 Mb.
Unlike in the first and second peaks, for the third peak 
the $^2D$ and $^2F$ states contribute significantly, while the $^2G$ state contributes 
moderately to the cross section.
The most important final states are

$4d^{10} 5s^2 5p^5 4f (^3G)\epsilon d (^2D)$,

$4d^{10} 5s 5p^6 4f (^3F)\epsilon p (^2D)$,

$4d^{10} 5s^2 5p^5 4f (^3F)\epsilon d (^2F)$,

$4d^{10} 5s^2 5p^5 4f (^3G)\epsilon d (^2F)$ 

$4d^{10} 5s 5p^6 4f (^3F)\epsilon p (^2F)$,

$4d^{10} 5s^2 5p^6 (^1S)\epsilon g (^2G)$.

From the above analysis we can see that the $^3D$, $^3F$ and $^3G$ terms of the 
core wave function
are more important than the $^1D$, $^1F$ and $^1G$ terms. 

Figure 2 displays the comparison of our RPAE results with the experimental data.
Solid and dotted curves represent, respectively our calculations and the data from the 
measurement [10]. 
The agreement is reasonable. The two main peaks in the experiment
are closer to each other than those in our calculation. Both the theoretical and experimental 
results are much larger than the cross section for photoionization of Ce@C$_{82}^+$
(see section a of Figure 2 in
Ref. [10]). The photoionization cross section for Ce@C$_{82}^+$ has a maximum of 20 Mb
around 123 eV. This value is much smaller than that of our third peak 57.0 Mb at 123.2 eV. 
The reduction effect from the carbon cage can also be seen from comparison of 
our results with the experimental data of the endohedral fullerene Ce@C$_{82}$ [7]. 
The photoionization cross section of the $4d-4f$ giant dipole resonance in Ce@C$_{82}$ 
is estimated to be 14.3 Mb at 130 eV, which is
also smaller than our results of 20.9 Mb at 130.0 eV.  
Therefore our calculation
further confirms the suppression effect obtained by the experiment 
when an atom encapsulated inside the  
carbon cage is photoionized. 

Many theoretical calculations
demonstrate confinement resonances [1-6]. However, all the experiments show no 
confinement resonances.
The measurements demonstrate a great suppression of the photoionization cross section
by the carbon cage.  
It is hypothesized in Ref. [10] that additional decay channels for the
Ce $4d$ vacancy may exist. After analyzing carefully the theoretical models 
and the experimental
arrangement,  we are led to the conclusion that the discrepancy between the 
theory and measurement
may be due to the assumed location of the atom inside the C$_{60}$ by theory. 
The location of the atom in the calculation 
of the confinement resonances is always assumed to be at the center of the carbon cage. 
The photoelectron
ionized from the atom will be reflected by the carbon sphere. For the atom assumed 
to be at the 
center of the C$_{60}$ the reflected wave and the incoming wave easily interfere with
each other and create the resonance. However,
the atoms in the current experiments involving the photoionization of 
Ce@C$_{82}^+$, Ce@C$_{82}$, Pr@C$_{82}$ etc.
are located at off-center positions and adjacent to the carbon cage.
This geometrical configuration causes the photoelectron ionized from for example, 
the Ce$^{3+}$ ion
to be multiply reflected to different directions by the carbon cage. The reflected 
wave and incoming wave do not necessarily readily interfere with each other, but are easily 
absorbed by the carbon cage in the process of photoionization. This may be the 
reason for the great suppression observed in the experiment.
 
To resolve directly the long-standing discrepancy between the measurements on the one hand
and theoretical
predictions on the other on the photoionization of an endohedral fullerene, we 
recommend that both measurements and calculations be performed on the 
photoionization of A@C$_{60}$ (atom at the center of C$_{60}$ cage).

\leftline{\bf Conclusion}

In conclusion, we have performed a RPAE calculation for the 
photoionization of the Ce$^{3+}$ - Ce$^{4+}$ process.  The reasonable agreement in 
magnitude and shape with the recent measurement [10] confirms the suppression effect
of the carbon cage in a endohedral fullerene. 
The plausible reasons for the confinement resonance or suppression
effect have been advanced and discussed as well.

\bigskip
\leftline{\bf Acknowledgments}
\bigskip

This work was supported by the U.S. DOE, Division of Chemical Sciences,
Geosciences and Biosciences, Office of Basic Energy Sciences, OER. 

Authors would like to thank Professor R. A. Phaneuf for providing 
them with their recently measured data for the Ce$^{3+}$ - Ce$^{4+}$ 
photoionization process as well as Dr. A. Baltenkov for valuable discussions.

\bigskip

\leftline{\bf Appendix: Coulomb matrix elements for the intershell coupling between Ce$^{3+}$}
\leftline{\bf $5s-\epsilon p$, $5p- \epsilon s, d$ and $4d-4f$ transitions}
\bigskip

In the following equations
$l_1^{n_1}$ is an open shell and the other $l^n$ subshells are closed shells.
$G_{L_1S_1}^{L_1' S_1'}$ etc. is the fractional parentage coefficient.

(1) For the Coulomb interaction
\begin{eqnarray}
&&|l_3^{n_3-1}[L_3'S_3']l_2^{n_2}[L_2 S_2] l_1^{n_1+1}[L_1' S_1'] L" S">
\rightarrow  \nonumber \\
&&|l_3^{n_3}[L_3 S_3]l_2^{n_2-1}[L_2' S_2'] l_1^{n_1}[L_1 S_1] [L_c' S_c'] l_6 L'S'> \nonumber
\end{eqnarray}

the Coulomb matrix element of time forward type is:
\begin{eqnarray}
&&\sum_{k} G_{L_1 S_1}^{L_1' S_1'}
{[L_c',S_c',L_1',S_1']^{1/2} [L'] \sqrt{n_1+1} \over  [S']}
\left\{\begin{array}{ccc}
l_1 & l_3 &  k \\
L" &  L  &  L_1'
\end{array}
\right \}
\left\{\begin{array}{ccc}
l_6 & l_2 & k \\
L &  L'  &   L_c'
\end{array} \right \} \nonumber \\
&*& <l_6||C^k||l_2>< l_3||C^k||l_1>R_k(l_6 l_3;l_2l_1)
(-1)^{L"+k+L+l_6+L_1'+L_c'+l_2+1}    \nonumber \\
&-&\sum_{k L_{16}} G_{L_1 S_1}^{L_1' S_1'} 
[S_1',L_1',L_c',S_c']^{1/2} [L', L_{16}] \sqrt{n_1+1} 
\left\{\begin{array}{ccc}
l_2 & L_1 &  L_c' \\
l_6 &  L'  &  L_{16} 
\end{array}
\right \}
\left\{\begin{array}{ccc}
l_2 & L_{16} & L' \\
L_1' &  l_3  &  k
\end{array} \right \} \nonumber \\
&& \left\{\begin{array}{ccc}
l_6 & l_1&  k \\
L_1' &  L_{16} &  L_1 
\end{array}
\right \}
\left\{\begin{array}{ccc}
1/2 & S_1 & S_c' \\
1/2 & S'  &  S_1'
\end{array} \right \} \nonumber \\
&*& <l_6||C^k||l_1>< l_3||C^k||l_2>R_k(l_6 l_3;l_2l_1)
(-1)^{2S_1'+l_1+l_3}   \nonumber \\
\end{eqnarray}

The exchange part of the Coulomb matrix element of time backward type is:
\begin{eqnarray}
&&-\sum_{k} G_{L_1S_1}^{L_1' S_1'}
[L_c", S_c"][L_1', S_1', L", L',L_c',S_c']^{1/2} \sqrt{n_1+1} \nonumber \\ 
&*& \left\{\begin{array}{ccc}
L' & L_c' &  l_6 \\
L_c" &  L  &  l_3 \\
l_1 & l_2 & k
\end{array}
\right \}
\left\{\begin{array}{ccc}
S_c' & 1/2  & S' \\
S_c" & 1/2  & S
\end{array} \right \}
\left\{\begin{array}{ccc}
S_1' & S"  & 1/2 \\
S_c" & S_1  & 1/2
\end{array} \right \}
\left\{\begin{array}{ccc}
L_1' & L" & l_3 \\
L_c" & L_1 & l_1
\end{array} \right \}                \nonumber \\
&*& <l_6||C^k||l_3><l_1||C^k||l_2>R_k(l_1 l_6;l_2 l_3)
(-1)^{2S_c"+1+n_1+L"+l_6+l_2-L_1} \nonumber
\end{eqnarray}
where $L_c" S_c"$ are from coupling of $l_2^{n_2-1}[L_2' S_2'] l_1^{n_1}[L_1 S_1] [L_c'S_c']$.
The symbols () and \{ \} are the so-called 3j symbol and 6j symbol, respectively. The symbol 
\{ \}
with three lines and three columns is the so-called 9j symbol. $C^k$ is the 
normalized spherical harmonics,
\begin{equation}
<l_3||C^k||l_1>=(-1)^{l_3}[l_3,l_1]^{1/2}\left(\begin{array}{ccc}
l_1 & k & l_3 \\
L' &  L_c'  &   L
\end{array} \right ) 
\end{equation}
and 
\begin{equation}
R_k(l_1 l_6;l_2 l_3)=\int R_{l_1}^* R_{l_6}^* {r^k_< \over r^{k+1}_> } R_{l_2} R_{l_3} r^2 dr  
\end{equation}

\leftline{\bf References}

\begin{enumerate}
\item M. J. Puska and R. M. Nieminen,
Phys. Rev. A {\bf 47}, 1181 (1993); {\bf 49}, 629 (1994).
\item M. E. Madjet, H. S. Chakraborty, and S. T. Manson,
Phys. Rev. Lett. {\bf 99}, 243003 (2007).
\item M. Ya. Amusia, A. S. Baltenkov, L. V. Chernysheva, Z. Felfli and A. Z. Msezane,
J. Phys. B {\bf 38}, L169 (2005).
\item M. Ya. Amusia, A. S. Baltenkov, and U. Becker,
Phys. Rev. A {\bf 62}, 012701 (2000).
\item V. K. Dolmatov and S. T. Manson,
J. Phys. B: At. Mol. Opt. Phys. {\bf 41}, 165001 (2008).
\item Zhifan Chen and Alfred Z. Msezane.
arXiv:0811.2185
\item K. Mitsuke, T. Mori, J. Kou, Y. Haruyama, and Y. Kubozono,            
J. of Chem. Phys. {\bf 122}, 064304 (2005).
\item H. Katayanagi {\it et al.}
J. Quant. Spectrosc. Radiat. Tranisfer {\bf 109}, 1590 (2008).             
\item A. M\"uller {\it et al.}
J. Phys. Conf. Ser. {\bf 88}, 012038 (2007).                              
\item A. M\"uller {\it et al.}
Phys. Rev. Lett. {\bf 101}, 133001 (2008).                               
\item K. Mitsuke {\it et al.} 
Int. J. Mass Spectrom. {\bf 243}, 121 (2005).    
\item Junqi Ding, Lu-Tao Weng, and Shihe Yang,
J. Phys. Chem. {\bf 100}, 11120 (1996)
\item Bing-Bing Liu {\it et al.},
Journal of Physics and Chemistry of Solid {\bf 58}, 1873 (1997).
\item K. Muthukumar and J. A. Larsson,
J. Phys. Chem. {\bf 112}, 1071 (2008).
\item Zhifan Chen and Alfred Z. Msezane,
J. Phys. B:At. Mol.Opt.Phys. {\bf 39}, 4355 (2006).
\item Zhifan Chen and Alfred Z. Msezane,
arXiv:0811.2715  
\item Zhifan Chen and Alfred Z. Msezane,
Phys. Rev. A {\bf 77}, 042703 (2008).
\item Giant Resonances in Atoms, Molecules, and Solid, NANO ASI series,
edited by J. P. Connerade, J. M. Esteve, and R. C. Karnatak,
(Plenum, New York, 1987).
\item Zhifan Chen and Alfred Z. Msezane,
Phys. Rev. A {\bf 72}, 050702(R) (2005).

\end{enumerate}

\leftline{\bf Figure Captions}
Fig. 1. The cross sections for the Ce$^{3+}$-Ce$^{4+}$ photoionization process
in the energy region 100-150 eV calculated in the RPAE approximation.
Dotted, dashed and dash-dotted curves represent, respectively the partial cross sections
for the symmetries $^2F$, $^2D$ and $^2G$. The solid curve is the total
photoionization cross section, which equals the sum of these partial 
cross sections. 

Fig. 2. Comparison of our RPAE results with the experimental data for the 
Ce$^{3+}$ - Ce$^{4+}$ photoionization process [10]. Solid and dotted curves are respectively, 
the calculated and measured results.

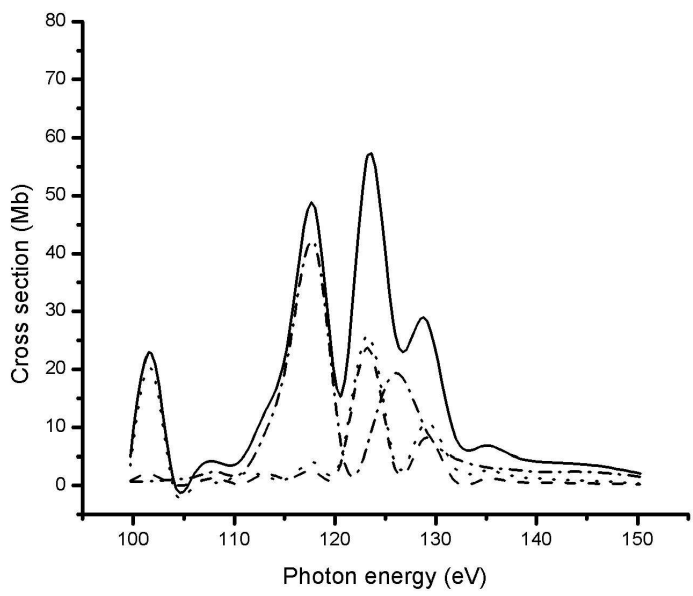
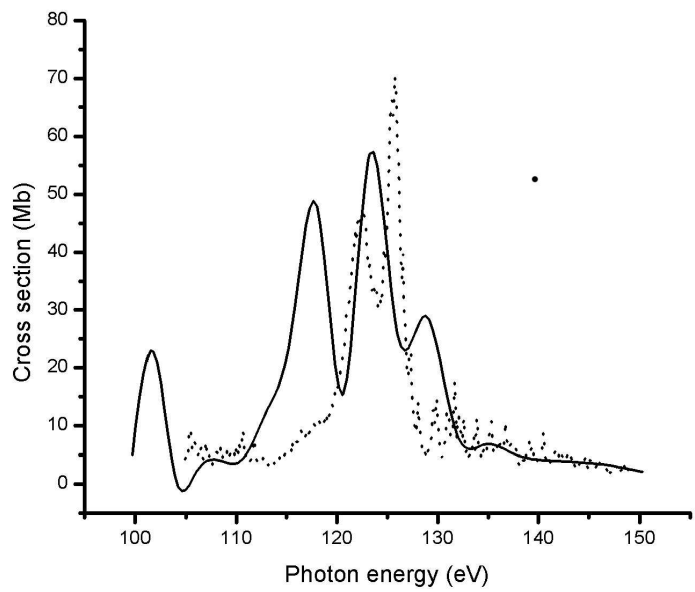
\end{document}